\documentclass[12pt]{article}
\usepackage{amssymb,amsfonts}
\usepackage{amsmath}
\begin{document}
\ \vskip0.5cm \centerline{\bf {GENERALIZED DARBOUX TRANSFORM }}
\centerline{\bf {IN THE ISHIMORI MAGNET MODEL }}
 \centerline {{\bf  ON THE BACKGROUND
OF SPIRAL STRUCTURES {\footnote {JETP LETTERS, Vol.78, No11, 2003,
pp.740-744.}}}}
 \vskip0.3cm {\centerline {\bf E.Sh.Gutshabash}}
\vskip0.3cm \centerline{\small {Research Institute of Physics,
St.Petersburg State University, St.Petersburg, 198504 Russia}}
\centerline{\small {e-mail: gutshab@EG2097.spb.edu}} \vskip0.3cm

\vskip0.5cm {\small The integration procedure based on the the
generalized Darboux transform is suggested for the Ishimori magnet
model. Exact solutions are constructed for the model of background
of spiral structures. The possibility of phase transition in the
system is hypothesized.}  \vskip0.4cm PACS: 52.35.Sb \vskip0.7cm

 It is well known that the phenological approach suggested by Landau and
 Lifshitz in the theory of ferromagnetism is based on the idea
 that the evolution of weakly excited states of a spin system in
 the long-wavelength limit can be described in terms of a constant-length
 magnetization vector (magnetic moment density) and is
 characterized by a certain effective field [1]. With allowance
 for the fact that the greatest contribution to the process comes
 from the exchange interaction between crystals atoms, this made it
 possible, in particular, to obtain nonlinear equation for the
 one-dimensional case (isotropic Heisenberg ferromagnet model),
 which was subsequently solved by the inverse scattering transform
 method (ISM)[2, 3]: exact solitonic solutions were obtained, excitation spectrum was
 described, and infinite set of conservation laws was found.
 Further progress in this direction was achieved in work
 [4], where the Lax representation was applied to the anisotropic
 Landau-Lifshitz model (and, thereby, it was proved to belong to
 the class of completely integrable models), and in [5], where its
 exact solutions were found by the "dressing" method. Noteworthy
 is also work [6], which was devoted to the construction of the
 integrable deformations of Heisenberg model.

The situation is much more complicated in the two-dimensional
case. The corresponding nonstationary equation for a
Landau-Lifshitz ferromagnet, both isotropic and anisotropic,
proves to be nonintegrable (see, e.g. [7]) and has exact solutions
only for rather specific cases. At the same time, there is much
evidence, both experimental [8] and obtained by numerical
simulation [9], of the existence of stable localized
two-dimensional excitations with a finite energy. For a
two-dimensional system, the spectrum of these excitations becomes
more diversified; in particular, new nontrivial topological
objects appear in the spectrum.

The model suggested by Ishimori in [10] is presently the major
tool for the phenomenological description of ferromagnets of
dimensionality (2+1). This is, primarily, due to the key property
of this model: it is completely integrable and allows the use of
ISM and ${\bar \partial}$-dressing procedure to construct a rather
broad class of solutions (vortices (lumps), rational exponential
solutions, instantons, etc.) on the trivial background [11-14],
while, by using a nonstandard Darboux transform [15], a physically
interesting solution can be obtained in the form of a vortex
circulating on circumference with a constant angular velocity.

However, it is worthy of note that the exact two-dimensional
solutions were found at the expense of necessity of introducing
nonlocal (along with the exchange) spin interactions into the
model. The physical mechanism of this interaction is as yet
unclear. In this connection, it should be emphasized that the
standard Heisenberg (exchange) interaction mechanism in the
so-called Schwinger-boson mean-field theory has become more
understandable only a relatively short time ago (see, e.g., [8]
and references cited therein). This gives grounds to hope that the
nature of nonlocality that provides a broad spectrum of reasonable
solutions and the corresponding observed physical objects will be
clarified in the future and that the Ishimori model is quite
realistic (within the framework of the adopted macroscopic
approach).

In this work, a new and rather effective integration procedure is
proposed for the Ishimori ferromagnet model. It opens the way for
constructing exact solutions, including the ones on the nontrivial
background. As to the ISM (or the ${\bar \partial}$-dressing
method), this may encounter considerable technical difficulties.

The Ishimori magnet model is given by

$$
{\bf S}_t={\bf S}\wedge({\bf S}_{xx}+\alpha^2{\bf S}_{yy})+u_y{\bf
S}_x+ u_x{\bf S}_y, \eqno(1a)
$$
$$
u_{xx}-\alpha^2u_{yy}=-2\alpha^2{\bf S} ({\bf S}_x \wedge {\bf
S}_y), \eqno(1b)
$$
where ${\bf S}(x,y,t)=(S_1, S_2, S_3)$ is the three-dimensional
magnetization vector, $|{\bf S}|=1,\:u=u(x,y,t)$ is the auxiliary
scalar real field, and the parameter $\alpha^2$ equals $\pm 1$.
The case $\alpha^2=1$ will be called Ishimori-I magnet model
(MI-I), and $\alpha^2=-1$ will be called MI-II.

Note that in the static limit (${\bf S}={\bf S}(x, y))$ and
$u={\mathrm const}$, the MI-I model transforms into the model of
two-dimensional isotropic Heisenberg ferromagnet (elliptic version
of the nonlinear $O(3)$ - sigma-model), which was integrated in
[16-17] by using the ISM with boundary conditions of the
spiral-structure type.

A characteristic feature of model (1) is the presence of
topological charge

$$
Q_{T}=\frac
{1}{4\pi}\int_{-\infty}^{\infty}\int_{-\infty}^{\infty} {\bf
S}({\bf S}_x\wedge {\bf S}_y) dx dy,  \eqno(2)
$$
which is conserved in the course of system evolution (integral of
motion) and represents the mapping of a unit 2-sphere onto the
2-sphere: $\tilde S^2\rightarrow \tilde S^2$. This mapping is
known to be characterized be the homotopic group $\pi_2(\tilde
S^2)=Z$, where $Z$ is the integer group, signifying that $Q_T$
must be integer. According to Eq.(1b), the scalar function
$u=u(x,t)$ is related to the topological charge production density
{\footnote {Although, strictly speaking, the function $u$ has no
direct physical meaning, the functions $u_y$ and $\:u_x$, related
to each other by Eq.(1b), can likely be interpreted as "frictions
coefficients" inducing forced precession of the magnetization
vector ${\bf S}$ along the the $x$ and $y$ axes, respectively.}.

The set of Eqs.(1) is integrated using the following associated
linear system:

$$
\Psi_y=\frac {1}{\alpha}S\Psi_x  ,\eqno(3a)
$$
$$
\Psi_t=-2iS\Psi_{xx}+Q\Psi_x  ,\eqno(3b)
$$
where $Q=u_yI+\alpha^3u_xS+i\alpha S_yS-iS_x, \:
\Psi=\Psi(x,y,t)\in Mat(2,\mathbb{C}),  \:S=\sum_{i=1}^3
S_i\sigma_i,\: \sigma_i$ are the standard Pauli matrices, and $I$
is a unit $2\times 2$ matrix. From its definition, the matrix $S$
has the following properties: $S=S^{\ast},\;S^2=I,\;\det S=-1$ and
${\mathrm Sp}\:S=0$ (the symbol ($\ast$) denotes Hermitian
conjugation).

In what follows, we restrict ourselves to the MI-II case
($\alpha=i$); the situation with MI-I can be analyzed in a similar
manner.

 We will solve Eq.(1) by the method of generalized Darboux matrix transform {\footnote {In
 [18], the approach based on the Darboux transform was applied to
 a system that is gauge equivalent to (3), and the dressing
 relations were obtained thereafter.}.} We also demand that system (3) be covariant
 ($U \to \tilde U,\:\Psi \to \tilde \Psi$) about the transformation of the form
 ($U \equiv S$){\footnote {In the literature, the scalar variant of this transformation
 is sometimes called Moutard transform [19].}}

$$
\tilde \Psi=\Omega(\Psi,\Psi_1)\Psi_1^{-1}, \eqno(4)
$$
where $\Psi_1=\Psi_1(x,y,t)$ is a certain nongenerate bare
solution to the system of Eqs.(3), and $\Omega(\Psi,\Psi_1)\equiv
\Omega(x,y,t) \in Mat(2,\mathbb {C})$ is a functional defined on
the pair set of matrix functions {\footnote {It follows from
Eq.(3a) that the matrix solutions $\Psi$ and $\Psi_1$ are related
to each other by the nonlinear relationship
$\Psi_y(\Psi_x)^{-1}=\Psi_{1y}(\Psi_{1x})^{-1}$.}.} One then
obtains from Eq.(3a) two dressing relations:

$$
\tilde U=\Omega \Psi_1 U \Psi_1^{-1} \Omega^{-1},\:\:\:\tilde
U=i\Omega_y (\Omega_x)^{-1}. \eqno(5)
$$

Hence, setting $z=x+iy,\:{\bar z}=x-iy,\:\partial_z=(1/2)(
\partial_x-i\partial_y),\: \partial_{\bar z}=(1/2)(
\partial_x+i\partial_y),$ and $\:W^{(0)}=\Psi_1^{-1}U\Psi_1$, we
obtain the first equation for the matrix $\Omega$ (${\mathrm
det}\:(I \pm U)={\mathrm det}\:(I \pm W^{(0)})=0$):

$$ (I-W^{(0)})(\Omega^{-1})_{\bar z}=0. \eqno(6) $$  Note that, due
to the symmetry relation ${\bar U}=-\sigma_2U\sigma_2$, the
following involutions are valid:

$$ \Psi=\sigma_2{\bar \Psi}\sigma_2,\:\:Q=\sigma_2{\bar
Q}\sigma_2,\:\:\Omega=\sigma_2{\bar \Omega}\sigma_2, \eqno(7) $$
which means, in particular, that Eq.(6) can be rewritten in the
"conjugate" form

$$ (I+W^{(0)})(\Omega^{-1})_z=0. \eqno(8) $$

To obtain the second equation for the matrix $\Omega$, one should
use Eq.(3b). This equation, however, is rather cumbersome. Taking
into account that Eqs.(5) and (6) can easily be expressed in terms
of function $\tilde \Psi$, the following consideration can best be
carried out for this function.

Using the identity $Q+UQ=-2i(I+U)(uI+U)_{\bar z},$ one obtain from
Eq.(3b):

$$
(I+U)\{\Psi_t+2i\Psi_{xx}+2i(u_{\bar z}I+U_{\bar
z})\Psi_x\}=0.\eqno(9)
$$
Let us transform this equation. Making allowance for the relation
$\Psi_{y \bar z}= - U_{\bar z}\Psi_x-iU\Psi_{x {\bar z}}$ that
follows from (3b), one has

$$
(I+U)\{\Psi_t+2i\Psi_{xx}+2iu_{\bar z}\Psi_x-2\Psi_{y\bar
z}-2iU\Psi_{x\bar z}\}=0. \eqno(10)
$$
After multiplying both sides of this equation by $U$ on the left
and adding together the resulting expression and Eq.(10), one
finds:

$$
(I+U)\{\Psi_t+2i\Psi_{zz}+2i\Psi_{{\bar z}{\bar z}}+4iu_{\bar
z}\Psi_{\bar z}\}=0.\eqno(11)
$$
Thus, after applying transformation (4) together with the
covariance requirement, one obtains two (and two analogous
conjugate) equations for the function $\tilde \Psi$:

$$
(I-\tilde U)\tilde \Psi_{\bar z}=0, \eqno(12)
$$
$$
(I+\tilde U)\{{\tilde \Psi}_t+2i{\tilde \Psi}_{zz}+2i{\tilde
\Psi}_{{\bar z}{\bar z}}+4i{\tilde u}_{\bar z}{\tilde \Psi}_{\bar
z}\}=0.\eqno(13)
$$
Next, taking into account the identities ${\bf S}({\bf S}_x \wedge
{\bf S}_y)=(1/(2i))\:{\mathrm Sp}\:(UU_xU_y)$ and

$$
{\mathrm Sp}\:({\tilde U}{\tilde U}_x{\tilde U}_y)={\mathrm
Sp}\:(UU_xU_y)+2i\triangle (\ln {\mathrm \det}\:{\tilde
\Psi})+2i{\mathrm Sp}\:\{[U,U_x]{\tilde \Psi}^{-1}{\tilde \Psi}_x+
$$
$$
\eqno(14)
$$
$$
+[U,U_y]\tilde \Psi^{-1} \tilde \Psi_y\}+4{\mathrm Sp}\:\{U[\tilde
\Psi^{-1}\tilde \Psi_y,\tilde \Psi^{-1}\tilde \Psi_x]\},
$$
where $\triangle$ is the two-dimensional Laplace operator, and
requiring that Eq.(1b) be covariant, we rewrite it as

$$
\triangle(\tilde u-u-2\:{\ln {\mathrm \det}\:\tilde
\Psi})=2\:{\mathrm Sp}\:\{[U,U_x]{\tilde \Psi}^{-1}{\tilde
\Psi}_x+[U,U_y]\tilde \Psi^{-1} \tilde \Psi_y-2iU[\tilde
\Psi^{-1}\tilde \Psi_y,\tilde \Psi^{-1}\tilde \Psi_x]\}. \eqno(15)
$$

The relevant expression for the dressed topological charge has the
form

$$
{\tilde Q}_T=Q_T+\frac {1}{4\pi}\int_{-\infty}^{+\infty}
\int_{-\infty}^{+\infty}\:dx\:dy\:\triangle \ln\:{\mathrm \det}\:
\tilde \Psi+
$$
$$
+\frac{1}{4\pi}\int_{-\infty}^{+\infty} \int_{-\infty}^{+\infty}
dx dy\: {\mathrm Sp}\:\ \{[U,U_x]{\tilde \Psi}^{-1}{\tilde
\Psi}_x+[U,U_y]\tilde \Psi^{-1} \tilde \Psi_y-2iU[\tilde
\Psi^{-1}\tilde \Psi_y,\tilde \Psi^{-1}\tilde \Psi_x]\}. \eqno(16)
$$

Introducing the notation
 $ \Psi[1]=\tilde
\Psi,\ldots, U[1]=\tilde U,\ldots, u[1]=\tilde u,\ldots,
Q_T[1]=\tilde Q_T,\ldots$, one can readily obtain from Eqs.(5),
(15), and (16) upon N-fold dressing of the starting bare solution
$U=U^{(1)}\equiv U[0]$ {\footnote {These formulas can also be
written in terms of the matrix functionals
$\Omega(\Psi_i,\:\Psi_j)$, where $\Psi_i,\:\Psi_j$ are certain
bare solutions to the set of Eqs.(3) with $S=S^{(1)}$.}} :

$$
U[N]=\left(\prod_{j=0}^{N-1}\Psi[N-j]\right)U\left(\prod_{j=0}^{N-1}\Psi[N-j]\right)^{-1},
\eqno(17)
$$
$$
u[N]=u+2\:\sum_{j=1}^N\ln {\mathrm \det}\:\Psi[j]+
\int_{-\infty}^{+\infty}
\int_{-\infty}^{+\infty}dx^{\prime}dy^{\prime}\:
G(x-x^{\prime},y-y^{\prime})\times
$$
$$
\eqno(18)
$$
$$
\times \sum_{j=1}^N{\mathrm Sp}\:A_j(x^{\prime},\:y^{\prime},\:t),
$$
$$
Q_T[N]=Q_T+\frac
{1}{4\pi}\int_{-\infty}^{+\infty}\int_{-\infty}^{+\infty}dx\:dy
\sum_{j=1}^N\triangle \ln {\mathrm \det}\:\Psi[j]+
$$
$$
\eqno (19)
$$
$$
+\frac{1}{4\pi}\int_{-\infty}^{+\infty}\int_{-\infty}^{+\infty}dx^{\prime}dy^{\prime}
\sum_{j=1}^N{\mathrm Sp}\:A_j(x^{\prime},\:y^{\prime},\:t),
$$
where  $A_j(x,\:y,\:t)=\Bigl [U[j-1],U_x[j-1]\Bigr ]{\tilde
\Psi[j]}^{-1}{\tilde \Psi}_x[j]+\Bigl [U[j-1],U_y[j-1]\Bigr
]\tilde \Psi[j]^{-1} \tilde \Psi_y[j]-2iU[j-1]\Bigl [\tilde
\Psi[j]^{-1}\tilde \Psi_y[j],\tilde \Psi[j]^{-1}\tilde
\Psi_x[j]\Bigr ],\:G(x,y)=(1/2\pi)\ln (x^2+y^2)$ is the Green's
function of the Laplace equation.

Turning back to the simple dressing, we note the Eqs.(12) and (13)
defines the whole collection of the solutions to model (1) (in
reflectionless ( in ISM terms) section of the problem). According
to Eqs.(12) and (13), four cases is possible:  1). $ \tilde
\Psi_{\bar z}=0$ and the braced expression in Eq.(13) is also
zero; 2). $\tilde \Psi_{\bar z}=0$ and the expression in braces is
nonzero; 3).  the situation opposite to (2); and 4). $ \tilde
\Psi_{\bar z}\ne 0$ and the expression in Eq.(13) is also nonzero
(i.e., the columns of this matrix belong to the kernels of the
degenerate $I-\tilde U$ и $I+\tilde U$ transformations,
respectively).

In this letter, we restrict ourselves only to the first case.
Then,

$$
\tilde \Psi_{\bar z}=0, \:\:\:\tilde {\Psi}_t+2i\tilde
\Psi_{zz}=0. \eqno(20)
$$
This system has the well-known polynomial solutions ($\tilde
\Psi=\{\tilde \Psi_{ij}\},i,j=1,2,\:{\tilde \Psi}_{22}=\bar
{\tilde \Psi}_{11},\:{\tilde \Psi}_{12}=-\bar {\tilde
\Psi}_{21}$)([10], [13]):
$$
\tilde \Psi_{11}(z,t)=\sum_{j=0}^{N_1}\sum_{m+2n=j}\frac{a_j}{m!
n!}(-\frac{1}{2} z)^m(-\frac{1}{2} it)^n,\:\:\: \tilde
\Psi_{21}(z,t)=\sum_{j=0}^{M_1}\sum_{m+2n=j}\frac{b_j}{m!
n!}(-\frac{1}{2} z)^m(-\frac{1}{2} it)^n, \eqno(21)
$$
where $N_1,\:M_1$ are natural numbers and $M_1=N_1-1,\:a_j$ and $
\:b_j$ are the complex numbers; and the first summation is over
all possible combinations of the numbers $m,\:n \geq 0$ such that
$m+2n=j$.

The bare solution to the set of Eqs.(1) is taken in the form of
vector function ${\bf S^{(1)}}=(0,\:\sin \Phi^{(1)},\:\cos
\Phi^{(1)})$, where $\Phi^{(1)}=\delta_0t+
\alpha_0x+\beta_0y+\gamma_0,\:\alpha_0,\:\beta_0,\:\gamma_0,\:\delta_0
\in \mathbb {R}$ are parameters), i.e., in the form of a
two-dimensional spiral structure with $Q_T=Q_T^{(1)}=0$ (according
to Eq.(2)). To determine the function
$u(x,\:y,\:t)=u^{(1)}(x,\:y,\:t)$ one should substitute this
vector into Eqs.(1a) and (1b). The requirement for the
compatibility of the resulting two linear equations gives, after
integration,
$$
u^{(1)}=g_0^{(1)}(y+\frac{\beta_0}{\alpha_0}x)+\int^s
g_1^{(1)}(y(s^{\prime})+\frac{\beta_0}{\alpha_0}x(s^{\prime}),t)\:ds^{\prime},
\eqno(22)
$$
where $g_0^{(1)}$ and $g_1^{(1)}$ are arbitrary functions. The
function $g_0^{(1)}$ is constant on the characteristic
$y+(\beta_0/\alpha_0)x={\mathrm const}$ and $s$ is its parameter.
Therefore, the explicit expression for $\tilde u$ is determined by
Eq.(18) with $N=1$ and by Eqs.(21) and (22).

Dressing relation (17) gives ($S_{+}=S_1+iS_2$)

$$
\tilde S_3(x,\:y,\:t)=\frac{\cos \Phi^{(1)}(|\tilde
\Psi_{11}|^2-|\tilde \Psi_{21}|^2)-i\sin \Phi^{(1)}(\tilde {\bar
\Psi}_{21} \tilde {\bar \Psi}_{11}-\tilde \Psi_{21}\tilde
\Psi_{11} )}{|\tilde \Psi_{11}|^2+|\tilde \Psi_{21}|^2},
$$
$$
\eqno(23)
$$
$$
\tilde S_{+}(x,\:y,\:t)=\frac{2\cos \Phi^{(1)} \tilde
\Psi_{21}\tilde {\bar \Psi}_{11}+i\sin \Phi^{(1)} (\tilde {\bar
\Psi}_{11}^2+\tilde {\bar \Psi}_{21}^2)} {|\tilde
\Psi_{11}|^2+|\tilde \Psi_{21}|^2}.
$$
Setting $\delta_0=0$ and $\:N_1=1$, we obtain the simplest static
anti(vortex) (one-lump) solution on the background of (also)
static spiral structure:

$$
\tilde S_3(x,y,t)=\frac{\cos \Phi^{(1)}
[|a_0-\frac{1}{2}a_1z|^2-|b_0|^2]+i\sin \Phi^{(1)}
[b_0(a_0-\frac{1}{2}a_1z)-{\bar b}_0({\bar a}_0-\frac{1}{2}{\bar
a}_1{\bar z})]}{|a_0-\frac{1}{2}a_1z|^2+|b_0|^2},
$$
$$
\eqno(24)
$$
$$
\tilde S_{+}(x,y,t)=\frac{2b_0\cos \Phi^{(1)}({\bar
a}_0-\frac{1}{2}{\bar a}_1{\bar z})+i\sin \Phi^{(1)}[b_0^2+({\bar
a}_0-\frac{1}{2}{\bar a}_1{\bar z})^2]}
{|a_0-\frac{1}{2}a_1z|^2+|b_0|^2}.
$$
The calculations in Eq.(16) show that $\tilde Q_T \to \infty$; the
divergence arises after the integration of the first two terms in
the braces. For $\delta_0=\alpha_0=\beta_0=\gamma_0=0$, solution
(24) transforms into a static (anti)vortex (on the trivial
background) with the topological charge ${\tilde Q}_T=-1$ (see
also [13]).

For $\delta_0 \neq 0$ and $N_1=2$, i.e. for $\tilde \Psi_{11}=a_0-
(a_1/2)z+(a_2/2)[(1/4)z^2-it]$ and $\:\tilde
\Psi_{21}=b_0-(1/2)b_1z$, formulas (23) describe the dynamic
(anti)two-vortex (two-lump) state on the background of (also)
dynamic spiral structure with $\tilde Q_T \to \infty$, and it
transforms to the state with $\tilde Q_T=-2$ if the parameters
entering $\Phi^{(1)}$ turn to zero.

Clearly, the intermediate types of solutions are also quite
realistic and of interest. Among these are a static vortex on the
background of the dynamic spiral structure and a dynamic vortex on
the background of the static spiral structure.

One the basis of these results, the hypothesis can be put forward
that the structural second-order spiral-vortex $\to $ vortex phase
transition (analogous to the Kosterlitz-Thouless transition) is
possible in the system of interest. This can occur if the
parameters $\delta_0,\:\alpha_0,\:\beta_0$ and $\gamma_0$ are
functions of time, i.e., functionals of an external nonstationary
and spatially uniform magnetic field {\footnote {One can show that
the addition of the term ${\bf S}\wedge {\bf H}(t)$ to the
right-hand side of Eq.(1a), where ${\bf H}(t)$ is the external
magnetic field, merely renormalizes the magnetization vector; i.e.
this term can be eliminated by the appropriate transformation
through making the bare solution dependent on magnetic field.}.}
Then the fact that the parameter turns to zero means that there is
a certain critical field ("Curie point" or, more precisely,
Lifshitz point) that corresponds to the phase transition point.
This hypothesis is confirmed by the experimental fact that the
spiral (modulated, incommensurate) structure in a magnetic field
can convert into the commensurate structure corresponding to a
paramagnet  with magnetic moments mainly oriented along the
external field {\footnote {Clearly, the theoretical justification
of this hypothesis should rest on the consideration of the order
parameter of system and on the analysis of the Ginzburg-Landau
functional in the vicinity of the critical point. However,
although the Hamiltonian of system (1) is known (generally
speaking, it is obtained in [21] for a modified MI model), the
relevant calculations become overly cumbersome even for the
simplest solutions and, thus, are beyond the scope of this
article.}}[20]. We also emphasize that this phase transition
should be accompanied by a change in a symmetry and topological
properties of the system.

Another series of solutions to model (1) can be found if the
solution to the system of Eqs.(20) is sought in the form

$$
\tilde \Psi_{11, 21}(x,\:y,\:t)=\int_{-\infty}^{\infty}
 B_{11, 21}(p)e^{-2ip^2
t+pz}dp. \eqno(25)
$$
Here $B_{11, 21}$ are the functional parameters. In particular, by
setting $B_{11, 21}(p)=c_{11, 21}\delta(p-p_{11, 21})$, where
$\delta(.)$ is the Dirac delta function and $c_{11, 21} \in
{\mathbb C}$ and $p_{11, 21} \in {\mathbb R}$ are parameters, one
gets using Eq.(17) ($c_{11, 21} \neq 0$ and the symbol c.c. stands
for the complex conjugation):

$$
{\tilde S}_3(x,y,t)=
$$
$$
=\frac {\cos
\Phi^{(1)}[|c_{11}|^2e^{2|p_{11}|x}-|c_{21}|^2e^{2|p_{21}|x}]-i\sin
\Phi^{(1)}[{\bar c_{11}}{\bar
c_{21}}e^{2i(p_{11}^2+p_{21}^2)t+2(p_{11}+p_{21}){\bar z}}-c.c.]}
{|c_{11}|^2e^{2p_{11}x}+|c_{21}|^2e^{2p_{21}x}},
$$
$$
\eqno (26)
$$
$$
{\tilde S}_{+}(x,\:y,\:t)=
$$
$$
=\frac {2{\bar c}_{11}c_{21}\cos
\Phi^{(1)}e^{2i(p_{11}^2-p_{21}^2)t+p_{11}{\bar z}+p_{21}z}+i\sin
\Phi^{(1)}({\bar c}_{11}^2e^{4ip_{11}^2t+2p_{11}{\bar
z}}+c_{21}^2e^{-4ip_{21}^2t+2p_{21}z})}
{|c_{11}|^2e^{2p_{11}x}+|c_{21}|^2e^{2p_{21}x}}.
$$
\vskip0.3cm Therefore, a exponential and nonsingular solution is
found on the background of spiral structure.  At $\Phi^{(1)} \to
0$, the solution is finite if $p_{11},\:p_{21}
> 0$; in this case the component $\tilde S_3$ evolves only along the
$x$ variable.

More complicated solutions of this type can be found if the
functionals $B_{11, 21}$ are taken as linear combinations of delta
functions.

Of interest is to compare the results obtained in this work with
the results of work [22], where the MI-II model was proved to be
gauge equivalent to the known hydrodynamic Davey-Stuartson-II
system (that describes the evolution of nearly monochromatic
small-amplitude wavepacket at the surface of a small depth fluid).
This means that the Lax pair can be transformed from one system to
another by a certain gauge transformation, which, in turn, allows
the formulas relating the solutions for these systems to be
derived. It is significant that the initial boundary-value
problems with specified classes should posses similar equivalence;
in [22], the class of rapidly decreasing Cauchy data was assumed
in both cases. Clearly, there is no gauge equivalence in the
considered case of spiral (and, hence, nondecreasing) structures,
while the solution constructed on their background cannot be
derived from the solution to the Davey-Stuartson-II model.

Note in conclusion that the approach developed in this work can
easily be extended to a series of Myrzakulov magnet models
[23,24], which are modifications of the Ishimori model; for them,
the first Lax-pair equation either is close or coincides with Eq.
(3a) and the main modifications concern the functional $Q$ in Eq.
(3b).

I am grateful to A.B.Borisov for attention to the work and
S.A.Zykov and A.V.Shirokov for assistance.

\vskip 2cm \centerline {{\bf \large {REFERENCES}}} \vskip 0.3cm

\vskip1.5cm   1. \parbox [t] {12.7cm}
       {{\em L.D.Landau,} - Collected Works (Gostekhizdat, Moscow, 1969), Vol.1,
       pp.128-143.}}

\vskip0.3cm   2. \parbox [t] {12.7cm}
       {{\em L.A.Takhtajan,} - Phys. Lett. \underline {64A}, 235 (1977).}

\vskip0.3cm   3. \parbox [t] {12.7cm}
       {{\em L.D.Faddeev and L.A.Takhtajan,} - Hamiltonian Methods in the Theory of Solitons
       (Nauka, Moscow, 1986; Springer, Berlin, 1987).}

\vskip0.3cm   4. \parbox [t] {12.7cm}
       {{\em E.K.Sklyanin,} - On complete integrability of the Landau-Lifshitz
       equation. Preprint No E-3-79. LOMI. Leningrad. 1979.}

\vskip0.3cm  5.  \parbox[t]{12.7cm}
       {{\em A.I.Bobenko,} - Zap. Nauchn. Semin. LOMI. \underline {123}. 58. (1983).}

\vskip0.3cm  6. \parbox [t] {12.7cm}
       {{\em A.V.Mikhailov and A.B.Shabat,} - Phys. Lett. \underline {116A}. 191. (1986).}

\vskip0.3cm  7. \parbox [t] {12.7cm}
       {{\em A.M.Kosevich, B.A.Ivanov and A.S.Kovalev,} - Phys. Rep. \underline {194}. 117.
        (1994).}

\vskip0.3cm  8. \parbox [t] {12.7cm}
       {{\em Tai Kai Ng,} - Phys. Rev. Lett. \underline {82}. 3504. (1999).}

\vskip0.3cm  9. \parbox [t] {12.7cm}
       {{\em B.Piette and W.J.Zakrzewski,} - Physica D. \underline {119}. 314. (1998).}

\vskip0.3cm   10. \parbox [t] {12.7cm}
       {{\em Y.Ishimori,} - Progr. Theor. Phys. \underline {72}. 33. (1984).}

 \vskip0.3cm 11. \parbox [t] {12.7cm}
       {{\em V.G.Dubrovsky and B.G.Konopelchenko,} - Coherent
       structures for the Ishimori equation. 1. Localized solitons
       with the stationary boundaries. Preprint No.90-76. Institute
       of Nuclear Physics. Novosibirsk. 1990.}

\vskip0.3cm  12. \parbox [t] {12.7cm}
       {{\em V.G.Dubrovsky and B.G.Konopelchenko,} - Coherent
       structures for the Ishimori equation. 2. Time-depend boundaries.
       Preprint No. 91-29. Institute
       of Nuclear Physics. Novosibirsk. 1990.}

\vskip0.3cm 13. \parbox [t] {12.7cm}
       {{\em B.G.Konopelchenko,} - Solitons in Multidimensions. World
       Scientific. (1993.)}

\vskip0.3cm 14. \parbox [t] {12.7cm}
       {{\em V.G.Mikhalev,} - Zap. Nauchn. Semin. LOMI {189}. 75. (1991).}

\vskip0.3cm 15. \parbox [t] {12.7cm}
       {{\em K.Imai and K.Nozaki,}- Progr. Theor. Phys. \underline {96}. 521. (1996).}

\vskip0.3cm 16. \parbox [t] {12.7cm}
       {{\em E.Sh.Gutshabash and V.D.Lipovskii,} - Teor.Mat.Fiz. \underline {90}. 175. (1992).}

\vskip0.3cm 17. \parbox [t] {12.7cm}
       {{\em G.G.Varzugin, E.Sh.Gutshabash and V.D.Lipovskii,} - Teor.Mat.Fiz.
       \underline {104.} 513. (1995).}

\vskip0.3cm 18. \parbox [t] {12.7cm}
       {{\em E.Sh.Gutshabash,} - Zap. Nauchn. Semin. Pomi., \underline {271}. 155. (2002);
        nlin.SI/0302002.}

\vskip0.3cm 19. \parbox [t] {12.7cm}
       {{\em C.Athorne and J.J.Nimmo,} - Inv. Prob. \underline {7}. 809. (1995).}

\vskip0.3cm 20. \parbox [t] {12.7cm}
        {{\em Yu.A.Izumov,} - Diffraction of neutrons on Long-Period Structures
        (Energoatomizdat, Moscow, 1987.)}

\vskip0.3cm 21. \parbox [t] {12.7cm}
       {{\em L.Martina, G.Profilo, G.Soliani and L.Solombrino,} - Phys. Rev. B.
       \underline {49}. 12915. (1994).}

\vskip0.3cm 22. \parbox [t] {12.7cm}
       {{\em V.D. Lipovskii and A.V.Shirokov,} - Funkts. Anal. Pril.
       \underline {23}. 65. (1989).}

 \vskip0.3cm 23. \parbox [t] {12.7cm}
        {{\em R.Myrzakulov,} - On some integrable and
        nonintegrable soliton equations of magnets. Preprint KSU.
        Alma-Ata. 1987.}

\vskip0.3cm 24. \parbox [t] {12.7cm}
        {{\em N.K.Bliev, G.Nugmanova, R.N.Syzdukova and
        R.Myrzakulov,} -
         Soliton equations in 2+1 dimensions: reductions,
        bilinearizations and simplest solutions. Preprint CNLP.
        No. 1997-05. Alma-Ata. 1997; solv/int 990214.}

\end {document}